\documentstyle[12pt]{article}
\newcommand{\gev}{\mbox{GeV}}
\newcommand{\gevc}{\mbox{$\gev /c$}}
\newcommand{\gevcc}{\mbox{$\gev /c^{2}$}}
\newcommand{\micron}{\mbox{$\mu$m}}
\newcommand{\cm}{\mbox{cm}}
\newcommand{\rz}{\mbox{$R-z$}}
\newcommand{\rphi}{\mbox{$R-\phi$}}
\newcommand{\Pt}[1]{\mbox{$P_{T}^{#1}$}}
\newcommand{\Ptmin}{\mbox{$P_{T}^{min}$}}

\newcommand{\lxy}{\mbox{$l_{xy}$}}
\newcommand{\fcorr}{\mbox{$F^{corr}$}}
\newcommand{\ctau}{\mbox{$c\tau$}}
\newcommand{\blife}{\mbox{$c\tau_{0}$}}
\newcommand{\intlum}{\mbox{$\int {\cal L} dt$}}

\newcommand{\alphab}{\mbox{$\alpha_{b}$}}
\newcommand{\fb}{\mbox{$f_{B}$}}
\newcommand{\fback}{\mbox{$f_{back}$}}
\newcommand{\fbkg}{\mbox{$B$}}
\newcommand{\fsig}{\mbox{$f_{sig}$}}
\newcommand{\normgaus}[2]{\mbox{$\frac{1}{\sqrt{2\pi}#2}$}
	\exp \left(-\frac{#1^{2}}{2#2^{2}}\right)}
\newcommand{\normexp}[2]{\mbox{$\frac{1}{#2}\exp \left(\frac{#1}{#2}\right)$}}
\newcommand{\psip}{\mbox{$\psi'$}}
\newcommand{\jpsi}{\mbox{$J/\psi$}}

\newcommand{\jpsipsip}{\jpsi,\psip}

\relax
\begin{document}

Contribution Code: GLS0208\\
Proposed Sessions: Pa-02,  Pl-19,  Pa-14,  Pl-15
\hfill Fermilab-Conf-94/136-E\\
\today

\vskip1cm

\begin{center}
\LARGE{\bf $\jpsipsip \to \mu^+\mu^-$ and $B
\to \jpsipsip$ Cross~Sections}
\vskip 2 cm
\normalsize
The CDF Collaboration
\vskip3cm

\vskip 3cm
Contributed paper to the 27th International Conference on \\
High Energy
Physics, Glasgow, July 20-27, 1994.

\end{center}
\vskip 1cm
\begin{abstract}
This paper presents a measurement of \jpsipsip\ differential cross
sections in $p\bar{p}$ collisions at $\sqrt{s} = 1.8~\mbox{TeV}$.  The
cross sections are measured above 4~\gevc\ in the central region
($|\eta| < 0.6$) using the dimuon decay channel.  The fraction of
events from $B$ decays is measured, and used to calculate $b$ quark
cross sections and direct \jpsipsip\ cross sections.  The direct cross
sections are found to be more than an order of magnitude
 above theoretical expectations.
\end{abstract}

\section{Introduction}

Charmonium production is currently the best way to study $b$ quark
production at CDF at the lowest possible transverse momentum
of the $b$.  While the signature and signal-to-noise for the \jpsi\
are excellent, the actual conclusions regarding $b$ production
are strongly dependent on the fraction of the data sample due to
$b$ decays.  One obvious way of determining this fraction is to
use the decay distance of the \jpsi\ state.
In contrast, the fraction of \psip's
from $b$ decays is thought to be close to one\cite{ref:glover}.
  Thus, in determining the \Pt{} spectrum
of $b$ quarks from a study of \psip\ events, one does not have to worry
about the fraction of \psip\ production due to $B$ decays.
In the course of this analysis, however, a very large
zero-lifetime component for the observed \psip\ signal has been
seen.  Therefore, one
still has to make use of the lifetime information.

\section{Detector Description}

The CDF detector has been described in detail
elsewhere\cite{ref:cdf_detect}.  We describe here briefly the
components relevant to this analysis.  The CDF coordinate system
defines the beam line to be  the $z$
direction.  $R$ is the radial distance from the beam
line, and $\phi$ is the azimuthal
angle.
A solenoidal magnet generating a 1.4 T magnetic field surrounds the
two tracking chambers used.
 The Central Tracking Chamber
(CTC) is a cylindrical drift chamber surrounding the beam line.

The CTC
contains 84 layers, which are grouped into nine superlayers, with 5
superlayers providing axial information and 4 providing stereo.
  It
covers a pseudorapidity range of $|\eta| < 1.1$.  The Silicon Vertex
Detector (SVX) is a silicon microvertex detector.  It consists of four
layers 2.9 to 7.9 \cm\ from the  beam line and
provides high resolution tracking information in the
\rphi\ plane.  When combined with a CTC track,
 it provides an impact parameter resolution of $(13 + 40/\Pt{})
\micron$ and a transverse momentum resolution of
$\sqrt{(0.0009\Pt{})^2 + (0.0066)^2}$.
 Primary vertices have a Gaussian distribution with a width
of about 27 \cm, while the SVX only reaches to $\pm 25 \cm$, so
 only about 60\% of the CTC tracks can be augmented with SVX
information.

Outside the CTC are electromagnetic and hadronic calorimeters, which
provide five absorption lengths of material before the Central Muon
Chambers (CMU).  The CMU consists of four layers of limited streamer
chambers.  It
is divided into 72 segments which cover 85\%
of the $\phi$ region for $|\eta| < 0.6$.
Muons with \Pt{} below $\sim 1.4~\gevc$  range out in the calorimeters.

The CDF trigger consists of three levels.  In the first level, both
muons must be detected in the CMU.  They must be seperated by at least 0.09
radians in $\phi$ and pass a slope cut in the $R-\phi$ plane.  The slope cut
is effectively a cut on the transverse momentum.  To pass the second
level, a fast hardware tracker must find at least one track
that points to the correct muon chamber.  For the third level, the full
CTC track reconstruction is done.  Both tracks must be found, and
extrapolate to within $4 \sigma$ of the muon
chamber tracks in $R-z$ and $R-\phi$.  $\sigma$ is the expected
multiple scattering error added in quadrature with the measurement
error.

\section{Event Selection}

Further cuts are applied offline to produce a purer sample.  The CTC
track and the CMU segment are required to match better than $3 \sigma$ in
\rphi, and $3.5 \sigma$ for the \jpsi\ and $3 \sigma$ for the \psip\
in \rz.  Hadronic energy is required in the calorimetry tower that
points to the muon chambers in the \psip\ events.  Both muons are
required to have $\Pt{\mu} > 2.0~\gevc$ and one muon must have
$\Pt{\mu} > 2.8~\gevc$. The dimuon is required to have $|\eta^{\mu\mu}| <
0.6$ and $\Pt{\mu\mu} > 4~\gevc$. Runs with known hardware problems were
excluded.  Because there are much better statistics in the \jpsi, a
tighter definition of bad runs was used there, resulting in a
smaller integrated luminosity.

For the \jpsi, the CTC track is beam constrained.  In the \psip, the
tracks are vertex constrained, using the SVX information if there are
three or more hits.  The $\chi^2$ of this fit is required to be
less than 10, with one degree of freedom.
The resulting invariant mass is used to define
a signal region of $3.0441 < m_{\mu\mu}<3.1443~\gevcc$ for the \jpsi\ and
 $3.636<m_{\mu\mu}<3.736~\gevcc$ for the \psip.
 Sideband regions
of $2.9606<m_{\mu\mu}<3.0274~\gevcc$ or $3.1610<m_{\mu\mu}<3.2278~\gevcc$ for
the \jpsi\ and
of $ 3.52<m_{\mu\mu}<3.62~\gevcc$ or $3.75<m_{\mu\mu}<3.85~\gevcc$ for
the \psip\ are also defined.

Figure~\ref{fig:mass} show the mass distribution of the events
passing these cuts.  There are a total of $26533 \pm 175$
\jpsi\ events and  $896 \pm 94$ \psip\ events.

\section{B Fraction}

For events where both muons have SVX tracks with at least three hits,
we vertex constrain the tracks to measure \lxy, the projection of the
decay length onto the \jpsipsip\ transverse momentum.  This is
converted into the proper lifetime of the parent by $\ctau = \lxy/[(\Pt{} /
m) \cdot \fcorr]$, where \fcorr\ is a Monte Carlo correction factor
that relates the \jpsipsip\ tranverse momentum to the transverse
momentum of the parent.

The background \ctau\ shape is measured from the sidebands.
It contains non-Gaussian tails, so we parametrize it by
$$
\fbkg(x) =
\left\{
\begin{array}{ll}
	(1-f_{+}-f_{-})\normgaus{x}{\sigma} +
		f_{+}\cdot \normexp{-x}{\lambda_{+}} &, x > 0 \\
	(1-f_{+}-f_{-})\normgaus{x}{\sigma} +
		f_{-}\cdot  \normexp{x}{\lambda_{-}} &, x \leq 0
\end{array}
\right.
$$
where
	$f_{+}$	is	the fraction in the positive exponential,
	$\lambda_{+}$ is the lifetime of the positive exponential,
	$f_{-}$	is	the fraction in the negative exponential,
	$\lambda_{-}$ is the lifetime of the negative exponential,
and	$\sigma$ is	the width of the central Gaussian.
The signal region is the fit to the function $N \cdot \fsig(\ctau)$, where
$$
\fsig(\ctau) = \fback\cdot\fbkg(\ctau) + (1-\fback) \left[(1-\fb)\cdot R(\ctau)
	+ \fb \cdot R\star E (\ctau)\right]
$$
where
$R(\ctau) =
f_r\normgaus{\ctau}{\sigma}+(1-f_r)\normexp{-|\ctau|}{\lambda}$ is the
resolution function and
$E (\ctau)  =  \normexp{-\ctau}{\blife}$.
\fback\	is	the background fraction,
\fb\	is	the B fraction,
\blife\	is	the B lifetime,
$\sigma$	is	the width of the Gaussian,
$f_r$	is	the fraction of non-Gaussian tails in resolution
function, and
$\lambda$ is the		lifetime of non-Gaussian tails.
 $R\star E (\ctau)$ is the convolution of the resolution function
with an exponential.
We fix \blife\ to 438 \micron, as found  by the CDF inclusive $B$
lifetime measurement.\cite{ref:incl_life}.  \fback\ is calculated by
$$
	\fback = \frac{\Delta m(Signal)}{\Delta m(Sideband)}
\cdot \frac{N_{Sideband}^{Entries}}{N_{Signal}^{Entries}} =
\frac{N_{Sideband}^{Entries}}{2 \cdot N_{Signal}^{Entries}}.
$$

In the \psip\, the data is fit using one unbinned log-likelihood fit.
  \fback\ is not
fixed, but a Poisson term is included for the likelihood that they
fitted value fluctuates to the calculated value.  $\sigma$ is
calculated on an event by event basis.
  Using the event-by-event error introduces non-Gaussian tails
into the total resolution function, so $f_r$ is fixed at 1.
The systematic error is estimated in part by allowing $f_r$ and
$\lambda$ to be fitted.
In the \jpsi, the signal and sidebands are fit in seperate binned
fits.
  \fback\ is fixed
and $\sigma$, $f_r$ and $\lambda$ are fit.
Figure~\ref{fig:life_fit_both}
show the result of the fits.

\subsection{\Pt{} dependence}

It is expected that the B fraction in the \jpsipsip\ samples rises
with \Pt{}.
To measure this, we divide the \psip\ sample into three \Pt{} bins
from $4-6~\gev$, $6-9~\gev$, and $9-20~\gev$.  The size of the last bin is
 neccesitated by the low statistics at high \Pt{}.
We then repeat the above fitting procedure in each of the \Pt{} regions.

In the \jpsi\ sample, the ``spectrum method'' is used to calculate the
\Pt{} dependence.  This is done in three steps.  First, the
\jpsi\ \Pt{} spectrum is obtained.  Then, the spectrum of events with
$\ctau > 250 \micron$ is found.  The two are normalized so that the
ratio of the total area is the inclusive $B$ fraction found above.
The $B$ fraction is then obtained by dividing the two distributions.
The results are shown in
Tables~\ref{tab:bfrac_jpsi}~and~\ref{tab:bfrac_psip}.

\begin{table}
\begin{center}
\begin{tabular}{||c|c||} \hline
\Pt{} (\gevc)	&	B Fraction (\%)
\\ \hline \hline
Above $4$	&	$19.6 \pm 1.5$ 	\\ \hline
$4-5 $		&	$13.3 \pm 1.0$ 	\\ \hline
$5-6 $		&	$15.9 \pm 1.2$ 	\\ \hline
$6-7 $		&	$21.0 \pm 1.7$ 	\\ \hline
$7-8 $		&	$25.2 \pm 2.1$ 	\\ \hline
$8-9 $		&	$25.2 \pm 2.2$ 	\\ \hline
$9-10 $		&	$26.9 \pm 2.5$ 	\\ \hline
$10-11 $	&	$34.3 \pm 3.3$ 	\\ \hline
$11-12 $	&	$31.0 \pm 3.5$ 	\\ \hline
$12-13 $	&	$32.4 \pm 4.1$ 	\\ \hline
$13-14 $	&	$42.1 \pm 5.7$ 	\\ \hline
$14-15 $	&	$27.6 \pm 5.0$ 	\\ \hline
\end{tabular}
\caption{Differential $B$ fraction and cross section (from all
sources) of \jpsi.  Errors are statistical plus systematic.}
\label{tab:bfrac_jpsi}
\end{center}
\end{table}
\begin{table}
\begin{center}
\begin{tabular}{||c|c|c||} \hline
\Pt{} (\gevc)	&	B Fraction (\%)
\\ \hline \hline

Above 4		&	$22.8 \pm 3.8$ 	\\ \hline
$4 - 6 $	&	$12.3 \pm 4.9$ 	\\ \hline
$6 - 9 $	&	$29.5 \pm 7.0$ 	\\ \hline
$9 - 20$	&	$39.3 \pm 7.9$ 	\\ \hline
\end{tabular}
\caption{Differential $B$ fraction and cross section (from all
sources) of \psip.  Errors are statistical plus systematic.}
\label{tab:bfrac_psip}
\end{center}
\end{table}

\section{Acceptance and Efficiencies}
\label{sec:accept}

The \jpsipsip\ differential cross section is defined by
$$
\frac{d\sigma}{d\Pt{}}(\Pt{}) =
	\frac{N}{\alpha \cdot \epsilon
		\cdot \intlum \cdot \Delta \Pt{}} \label{eq:xsec}
$$

where $N$ is number of \jpsipsip\ candidates in the  bin,
$\alpha$ is the detector and kinematic acceptance,
$\epsilon$ is the  trigger and
cut efficiency, \intlum\ is the integrated lumonosity, and
$\Delta \Pt{}$ is the size of the \Pt{} bin.
The acceptance was determined from several Monte Carlo data sets.  The
\psip\ acceptance used  \psip's generated flat in \Pt{} and
$\eta$.
The \jpsi\  acceptance  used several sets where $b$ quarks
were generated in different \Pt{} ranges, in order to provide
sufficient statistics in all regions.  The $b$ quarks were
generated using the next-to-leading order QCD
calculation\cite{ref:NDE} and MRSD0 structure
functions.  The quarks were fragmented to $B$ mesons
using Peterson fragmentation\cite{ref:peterson}
 with $\epsilon_b = 0.006 \pm 0.002$.  A
fast detector simulation was used on the events, and  the kinematic cuts
were then applied to the events.  The \jpsipsip\ acceptance is the
ratio of events that survived this process to the number of generated
events.

 The $b \to \jpsipsip$ acceptance  used events where $b$ quarks
were generated and forced to decay to \jpsipsip.  The \jpsipsip\ was
forced to have the momentum spectrum measured by
\cite{ref:cleo_thesis},\cite{ref:cleo_b_decays}.
The \Pt{} of the $b$ quark is described by \Ptmin.
\Ptmin\ is defined as the \Pt{} such that about 90\% of the b quarks
in our sample have $\Pt{b} > \Ptmin$.
The $b$ quark acceptance is then calculated by
$$
\alphab =
 \frac{N_{b}(\Pt{\jpsipsip} > 4.0~\gevc, |\eta^{\jpsipsip}| < 0.6,
	 |y^{b}| < 1.0)}
    {N_{b}(\Pt{b} > \Ptmin, |y^{b}| < 1.0)}
$$
The lack of a \Pt{} cut on the $b$ in the numerator corrects for the
fact that we are unable to remove those events with
$\Pt{b} < \Ptmin$.

The efficiencies of the three triggers where studied with unbiased \jpsi\
events, as were the muon quality cuts.
The efficiency of the CTC track reconstruction is $98.9 \pm
1.0\%$.
The efficiency of the muon reconstruction is $98 \pm
1\%$.

\section{Systematics}
\label{sec:syst}

Systematic errors can arise in three distinct areas: measuring the
\jpsipsip\ cross sections, determining what portion of that cross
section is from $B$ decays, and extrapolating those results to a $b$
quark cross section.

\subsection{\jpsipsip\ Cross Section Systematics}

We calculate the uncertainty due to the trigger
shape by varying the shape of the level 1 and
level 2 trigger efficiency parametrizations.  This is an 8\% effect in
the \jpsi\, and a 6.6\% effect in the \psip.
We estimate the effect of the acceptance correction in the \psip\ by
varying the acceptance by the statistical error and by
fitting the acceptance to a fifth degree polynomial, which is a 3.1\%
effect.  In the \jpsi,
fits to different order polynomials were used, giving a 1.5\% error.
A 5\% error is assigned due to the \jpsipsip\ polarization based on a
$\sim 10\%$ difference between the acceptance of maximallly polarized
\jpsi's with extreme values of $\alpha = \pm 1$.
The 1\% error measuring the cut efficiencies is neglibable compared to
other errors.
There is a 2\% error associated with each of the level 3 trigger
efficiency, the CTC tracking efficiency, and the muon reconstruction
efficiency.
We assign a 4\% systematic to the uncertainty in the luminosity.

\subsection{B Fraction Systematics}

Potential sources of systematic errors arise from the $B$ lifetime,
the spread of the $\sigma_{\ctau}$ distribution, and the fitting
technique.

A single value for the \ctau\ resolution is assumed in the \jpsi\ fit,
while a wide range exists.  The effect of this is estimated by
applying a cut of $\sigma_{\ctau} < 60 \micron$, which removes the long
tail in the distribution.  $\sigma_{\ctau}$ is related to the
opening angle between the muons, which decreases at higher \Pt{}.
Since the \jpsi\ from $B$ decays have a harder \Pt{} spectrum, this
may bias the $B$ fraction, which is lower with this cut.  The 7\%
difference is used as the systematic error.  In the \psip, the
unbinned fit uses the event-by-event error, so this effect is missing.
Varying the $B$ lifetime by one sigma changes the $B$ fraction by 0.002 or
less.  This is negligible compared to other contributions.

Changing the fitting technique and the form of the fitted function in
the \psip\ can change the $B$ fraction by 7\%, which is used as the
systematic error.

\subsection{B  Cross Section Systematics}

We assign a $1.3\%$ systematic error due to the uncertainty in the
rapidity spectrum in the range $0.6 < |y^{b}| < 1.0$.
We take an uncertainty of $4.6\%$
due to the uncertainty in the NDE spectrum with different values of
$\mu$ and $\Lambda$.
We assign 5\% based on half the difference in the acceptance calculated with
Peterson epsilon set to 0.004 and 0.008.
  A systematic of 5\% is assigned
based on the uncertainty in the CLEO \jpsipsip\ momentum distribution.
We assign a 1\% error because of the statistics in the b to \psip\
Monte Carlo.

In the \psip, the
effect of the large bin sizes for \fb\ is estimated by fitting the
B fraction to a straight line, and using a bin-by-bin correction factor.

\section{Cross Section}

\subsection{\jpsipsip\ Cross Section}

\newcommand{\showpsixsec}[5]{
$$
\sigma(p \bar{p} \rightarrow #1, \Pt{#1} < 4~\gevc, |\eta| < 0.6)
\cdot Br(#1 \rightarrow \mu^{+} \mu^{-}) =
$$
$$
\begin{array}{lcr}
 #2\pm #3(stat)\ #4 (syst)~\mbox{nb}&& \mbox{#5}
\end{array}
$$
}
\newcommand{\showjpsixsec}[5]{
\showpsixsec{\jpsi}{#1}{#2}{^{+#3}_{-#4}}{#5}
}
\newcommand{\showpsipxsec}[4]{
\showpsixsec{\psip}{#1}{#2}{\pm #3}{#4}
}
The cross sections are shown in Figure \ref{fig:XSec}.
The integrated cross sections are
\showjpsixsec{29.10}{0.19}{3.05}{2.84}{}
\showpsipxsec{0.721}{0.058}{0.072}{}

The cross section from $B$ decays is extracted by multiplying the
differential $B$ fraction by the cross section.  We can also obtain a
prompt cross section by multiplying the cross section by one minus the
$B$ fraction.  The results are shown in Figures~\ref{fig:XSec}
and~\ref{fig:xsec_jpsi}.  The prompt theory curves are from~\cite{ref:mlm}.

\subsection{Inclusive $b$ Cross Section}

We now convert these cross sections into an inclusive $b$ cross section.
We do so using the standard formula,
$$
\sigma(p \bar{p} \rightarrow bX, |\eta^{b}| < 1.0, \Pt{b} > \Ptmin) =
$$
$$
\frac{1}{2}\frac{1}{\intlum \cdot \alphab}
\frac{1}{Br(b\rightarrow \jpsipsip X) \cdot Br(\jpsipsip \rightarrow \mu\mu) }
{
	\left( \sum_{\Pt{}}
		\frac{N^{}(\Pt{})}
			{\alpha(\Pt{}) \cdot \epsilon(\Pt{})}
	\right)
}
$$
The factor of $\frac{1}{2}$ results from the fact that the $b$ or the
$\bar{b}$ can decay into the \jpsipsip.  We use branching ratios of
 $Br(b\rightarrow \jpsi X) = 1.16 \pm 0.09\%$
\cite{ref:cleo_lepton_photon_93},
 $Br(b\rightarrow \psip X) = 0.30 \pm 0.06 \%$\cite{ref:cleo_b_decays},
$Br(\jpsi \rightarrow \mu\mu) = 6.27 \pm 0.20 \%$\cite{ref:pdg_92} and
$Br(\psip \rightarrow \mu\mu) = 0.88 \pm 0.13 \%$\cite{ref:pdg_90}.
Our results are

\newcommand{\showbxsec}[4]{
$$
\begin{array}{lcr}
\sigma(p \bar{p} \rightarrow bX, |y^{b}| < 1.0,
\Pt{b} > #1~\gevc) =
 #2\pm #3\ \mu\mbox{b} & & \mbox{#4}
\end{array}
$$
}

\showbxsec{6.0}{12.16}{2.07}{\jpsi}
\showbxsec{7.3}{8.24}{1.34}{\jpsi}
\showbxsec{8.7}{5.20}{0.83}{\jpsi}
\showbxsec{5.9}{6.12}{2.04}{\psip}
\showbxsec{8.3}{3.85}{1.23}{\psip}

The error is statistical and systematic errors, added in quadrature.

\section{Acknowledgements}
     We thank the Fermilab staff and the technical staffs of the
participating institutions for their vital contributions.  This work
was
supported by the U.S. Department of Energy and National Science
Foundation;
the Italian Istituto Nazionale di Fisica Nucleare; the Ministry of
Education,
Science and Culture of Japan; the Natural Sciences and Engineering
Research
Council of Canada; the National Science Council of the Republic of
China;
the A. P. Sloan Foundation; and the Alexander von Humboldt-Stiftung.

\newpage
\newcommand{\etal}{{\em et al.}}


\begin{thebibliography}{99}

\bibitem{ref:glover} S.D. Ellis, \etal, {\sl Phys. Rev. Lett.} {\bf
36}, 1263 (1976); C.E. Carlson and R. Suaya, {\sl Phys. Rev. D} {\bf
15}, 1416 (1977)

\bibitem{ref:cdf_detect} F.Abe \etal, CDF Collab., {\sl Nucl.
Instrum. Methods Phys. Res.}, Sect. A, {\bf 271}, 387 (1988)

\bibitem{ref:NDE} P. Nason \etal, {\sl Nucl. Phys.} {\bf B327}
 49 (1988)

\bibitem{ref:incl_life} F. Abe \etal, {\sl Published Proceedings
Advanced Study Conference On Heavy Flavours}, Pavia, Italy, September
3-7, 1993.  FERMILAB-CONF-93/319-E

\bibitem{ref:cleo_thesis} W. Chen, {\sl Decays of Upsilon Mesons to
the \jpsi\ Meson and a Search for the Higgs Boson in $B$ Meson Decay},
Purdue University Thesis, May 1990

\bibitem{ref:cleo_b_decays} T. Browder \etal,
``A Review of Hadronic and Rare B Decays'', To appear in {\sl $B$ Decays},
2nd edition, Ed. by S. Stone, World Scientific

\bibitem{ref:cleo_lepton_photon_93} Y. Kubota \etal,
``Inclusive Decays of $B$ Mesons
to \jpsi\ and \psip,'' Contributed paper at XVI International
Symposium on Lepton-Photon Interactions: Cornell University, Ithaca,
N.Y., U.S.A., August                 10-15, 1993


\bibitem{ref:pdg_92} Particle Data Group, M.Aguillar-Benitez \etal,
{\sl Phys. Rev.} {\bf D45}, (1992)

\bibitem{ref:peterson} C. Peterson \etal, {\sl Phys. Rev. D}, {\bf
27}, 105 (1983)

\bibitem{ref:pdg_90} G.J. Feldman and M.L. Perl, {\sl Physics
Letters}, {\bf 33C}, 285 (1977)

\bibitem{ref:mlm} E. Braaten \etal, Fermilab Preprint,
FERMILAB-PUB-94/135-T (1994)
\end{thebibliography}
\end{document}